\documentclass[prd,twocolumn,showpacs,preprintnumbers,amsmath,amssy mb]{revtex4}

\usepackage{graphicx}
\newcommand{\beq}{\begin{equation}}
\newcommand{\eeq}{\end{equation}}

\newcommand{\bea}{\begin{eqnarray}}
\newcommand{\eea}{\end{eqnarray}}
\newcommand{\nn}{\nonumber}

\def\beq{\begin{equation}}
\def\eeq{\end{equation}}

\begin{document}

\title{Alfven modes driven nonlinearly by 
metric perturbations in\\ Anisotropic 
Magnetized Cosmologies}

\preprint{}

\author{Apostolos Kuiroukidis$^{1,2}$, Kostas Kleidis$^{1,3}$ and Demetrios B. Papadopoulos$^{1}$}


\affiliation{$^{1}$Department of Physics, Aristotle 
University of Thessaloniki, 54124 Thessaloniki, Greece\\
$^{2}$Department of Informatics, Technological Education 
Institute of Serres, 62124 Serres, Greece\\
$^{3}$Department of Civil Engineering, Technological Education 
Institute of Serres, 62124 Serres, Greece}

\date{\today}

\date{Received ...; accepted ...}


\begin{abstract}
We consider anisotropic magnetized cosmologies filled with 
conductive plasma 
fluid and study the implications of metric perturbations that 
propagate parallel to the ambient magnetic field. It is known that 
in the first order (linear) approximation with respect to the 
amplitude of the perturbations no electric field and density 
perturbations arise. However when we consider the non-linear coupling 
of the metric perturbations with their temporal derivatives, 
certain classes of solutions can induce steeply increasing in time, 
electric field perturbations. This is verified both 
numerically and analytically. The source of these perturbations can 
be either high-frequency quantum vacuum fluctuations, driven by the 
cosmological pump field, in the early stages of the evolution of the 
Universe, or astrophysical processes, or a non-linear isotropization 
process, of an initially anisotropic cosmological spacetime. 
\end{abstract}


\maketitle

\section{Introduction}

Magnetic fields are known to have a widespread presence in 
our Universe, being a common property of the intergalactic 
medium in galaxy clusters \cite{kron1}, while, reports on 
Faraday rotation imply significant magnetic fields in 
condensations at high redshifts \cite{kron2}. Studies of 
large-scale magnetic fields and their potential implications 
for the formation and the evolution of the observed structures, 
have been the subject of continuous investigation (see e.g. 
\cite{thorn}-\cite{jedam} for a representative though incomplete 
list). Magnetic fields observed in galaxies and galaxy clusters 
are in energy equipartition with the gas and the cosmic rays 
\cite{wolf}. The origin of these fields, whether of astrophysical 
or cosmological origin, remains an unresolved issue. 

If magnetism has a cosmological origin, as observations of $\mu G$ 
fields in galaxy clusters and high-redshift protogalaxies seem to 
suggest, it could have affected the evolution of the Universe 
\cite{giov}. There are several scenarios for the generation 
of primordial magnetic fields (see e.g. \cite{gras}). Most 
of the early treatments were Newtonian, while relativistic 
treatments appeared recently in the literature. A common factor in 
all these approaches is the MHD approximation, namely the 
assumption that the magnetic field lines are effectively frozen 
in an infinitely conducting cosmic medium 
(i.e., of zero resistivity). With few exceptions 
(e.g. \cite{fenn},\cite{jedam1},\cite{vlah}) the role of 
non-zero resistivity and kinetic viscosity have been 
ignored, these features however being essential for 
a comprehensive picture of the non-linear magnetic field 
evolution. The electric fields associated with the resistivity 
can be the source for particle acceleration, while the induced 
non-linear currents may react back upon the magnetic 
field \cite{vlah}. 

Many recent studies use a Newtonian or a FRW cosmological 
model to represent the Universe and super-impose a large-scale 
ordered magnetic field. The magnetic field is assumed to be 
too weak to destroy the FRW isotropy and the anisotropy, induced 
by it, is treated as a perturbation \cite{ruzm}, \cite{tsag}, 
\cite{durr}. Current 
observations provide a strong motivation for the adoption 
of a FRW model, but the uncertainities of the cosmological 
{\it Standard Model} are several. Therefore the approximation 
of neglecting the background anisotropy and magnetic fields, 
may lead to effects and phenomena that are absent in the 
above treatments. 
Within this context the formation of small-scale structures 
and the excitation of resistive instabilities in Bianchi-type 
models has been explored long ago \cite{fenn}, but the 
issue of excitation of MHD modes in anisotropic cosmological 
models and their subsequent temporal evolution is far from being 
exhausted \cite{pap2}.

Usually one assumes that $h^{2}$ is small. However one may 
consider strong GWs either in the Early Universe when 
dynamical isotropization of an anisotopic cosmological model 
takes place, or due to amplification of quantum vacuum fluctuations 
from the the variable gravitational field (cosmological pump field) 
of the Universe (metric tensor). Also $h\dot{h}$ can be large 
due to high frequency gravitational wave perturbations in 
the Early Universe. 

On the other hand examples of electrovacuum cylindically symmetric 
spacetimes, interacting with GWs, in the full theory are well 
known (\cite{pap4}, \cite{pap5}), or the extraction of GWs 
by black hole collisions \cite{abra} whereas the covariant 
decomposition into scalar, vector and tensor perturbations is 
treated in \cite{mukh}. Gravitational waves can carry a large 
amount of energy near the sources where they are generated 
(see e.g. \cite{kokk}). Though they do not interact much with 
matter under normal conditions, in the linear level, it has 
been shown that they can excite various kinds of plasma waves, 
more efficiently with increasing background magnetic field 
(see e.g. \cite{pap3}, \cite{brod1}, \cite{brod2}, 
\cite{serv1}, \cite{serv2}, \cite{moor1}, \cite{moor2}, 
\cite{kall}). Non-linear effects have been studied in 
\cite{brod1}, \cite{kall}.

However the problem of the behaviour of conductive plasmas, in 
anisotropic magnetised cosmologies, which are driven by external 
perturbations, in the full non-linear level is more or less 
open. Motivated by this fact we consider this problem in the 
background of Thorne's class of anisotropic magnetized cosmologies. 
We write the perturbed Einstein's equations in a manner that 
takes into account the non-linear coupling of the metric perturbations 
with their temporal derivatives. Then we study mainly the electric 
field perturbations induced by classes of solutions, to the 
evolution equations of the metric perturbations (which however 
are written in the linear approximation). The organization of 
this paper is as follows: In Section II we present the derivation 
of the closed set of equations used, aided by the two 
appendices. In Section III we present the various classes of 
analytical solutions that enter in our discussion, classified 
into small and large-t solutions. In Section IV we present 
the numerical results of our paper and finally in Section 
V we present a brief discussion of the results.

\section{Basic equations}

We consider classes of anisotropic magnetized cosmologies 
with mater content in the form of a perfect fluid and with 
the magnetic field in the $z$-direction. The metric is 
taken to be  
\bea
\label{metr}
(g_{TT})_{\mu \nu }=
\left(
\begin{array}{cccc}
-1,   &0,          &0,          &0\\
0,    &A^{2}+h_{+},&h_{\times},      &0\\
0,    &h_{\times},      &A^{2}-h_{+},&0\\
0,    &0,          &0,          &W^{2}\\
\end{array}\right)
\eea
where $A=A(t)$, $W=W(t)$,  
for the metric components of the background spacetime 
(see Appendix A for the class of Thorne's anisotropic 
magnetized cosmologies that is used), 
while the metric perturbations are assumed of the form 
$h_{+}=h_{+}(t,z)$ and $h_{\times}=h_{\times}(t,z)$.

The energy-momentum tensor for the perfect fluid is 
taken as $T^{(fl)}_{\mu \nu }=T_{\mu \nu }^{(0)}+
(\delta T_{\mu \nu })$ where of course in this expansion 
we use the background value\\ 
$T_{\mu \nu }^{(0)}=diag(\rho (t), A^{2}p(t), A^{2}p(t), 
W^{2}(t)p(t))$ and the perturbation is computed as 
\bea
\label{fluten}
(\delta T_{\mu \nu })=
\left(
\begin{array}{cccc}
\delta \rho,   &\delta T_{0x},&\delta T_{0y},&\delta T_{0z}\\
\delta T_{0x},    &A^{2}\delta p+ph_{+},&ph_{\times}, &0\\
\delta T_{0y},    &ph_{\times}, &A^{2}\delta p-ph_{+},&0\\
\delta T_{0z},    &0,          &0,          &W^{2}\delta p\\
\end{array}\right)
\eea
Here we have $\delta \rho =\delta \rho (t,z)$,  
$\delta p=\delta p(t,z)$, for the diagonal terms, 
$\delta T_{0x}=-(\rho +p)A^{2}(\delta u^{x})$, 
$\delta T_{0y}=-(\rho +p)A^{2}(\delta u^{y})$ and 
$\delta T_{0z}=-(\rho +p)W^{2}(\delta u^{z})$. Also we 
have used the four-velocity perturbations around the 
background (comoving) value $u^{\mu }_{(0)}=(1,0,0,0)$ 
and perturbation of the condition $u^{\mu }u_{\mu }=-1$ 
ensures $\delta u^{0}=0$. 

The energy-momntum tensor for the EM-field is constructed 
from the field tensor 
\bea
\label{emten}
F^{\mu \nu }=
\left(
\begin{array}{cccc}
0,    &E^{x}, &E^{y},   &E^{z}\\
-E^{x},    &0,&B^{z},   &-B^{y}\\
-E^{y},    &-B^{z}, &0, &B^{x}\\
-E^{z},    &B^{y},   &-B^{x}, &0\\
\end{array}\right)
\eea
where $E^{j}=F^{j\mu }u_{\mu }$, 
$B^{k}=(1/2)\epsilon ^{kbcd}u_{b}F_{cd}$ are 
the electric and magnetic field respectively, and the 
background magnetic field is assumed (as in Appendix A) 
to be $\vec{B}_{0}(t)=B_{0}(t)\hat{z}$. In general we 
assume the perturbations $\delta E^{j}=\delta E^{j}(t,z)$ 
and $\delta B^{j}=\delta B^{j}(t,z),\; \; (j=x,y,z)$ and 
the energy-momentum tensor is 
\bea
\label{emenerg} 
T^{\alpha \beta }_{(em)}&=&\frac{1}{4\pi }
[F^{\alpha \mu}F^{\beta \nu}g_{\mu \nu }-\frac{1}{4}g^{\alpha \beta }
F^{2}],\; \nn \\ 
(F^{2}&:=&F^{\alpha \mu }F^{\beta \nu}g_{\alpha \beta}g_{\mu \nu})
\eea

For the metric of Eq. (\ref{metr}) the components of the 
Einstein tensor are computed. Then these are expanded up to 
the second order and we set ($G=1=c$)
\bea
\label{expan} 
G_{\mu \nu }&:=&
G_{\mu \nu }^{(0)}+\delta G_{\mu \nu }
+{\cal O}(h^{3})=\nn \\
&=&8\pi (T_{\mu \nu }^{(0)}+\delta T_{\mu \nu })
\eea
Here on the r.h.s we have the total energy-momentum tensor,  
both form the fluid and the EM field, expanded 
with respect to the fluid and EM-field perturbations. The 
$\delta G_{\mu \nu }-$term contains in general 
first and second order corrections, 
with resect ot the GW amplitude, (though here the first 
order terms vanish as it is well known for the Alfven 
modes \cite{pap3})
and we set, as indicated below,\\ 
$\delta G_{\mu \nu }=8\pi \delta T_{\mu \nu }$. Also equating 
the zeroth order terms we obtain the usual Einstein's 
equation for the background 
\bea 
\label{einback}
\left(\frac{\dot{A}}{A}\right)^{2}+
2\left(\frac{\dot{A}\dot{W}}{AW}\right)&=&
8\pi \rho +A^{4}(B_{0})^{2}\nn \\
-\frac{\ddot{A}}{A}-\frac{\ddot{W}}{W}-
\left(\frac{\dot{A}\dot{W}}{AW}\right)&=&
8\pi p+A^{4}(B_{0})^{2}\nn \\
-2\frac{\ddot{A}}{A}-\left(\frac{\dot{A}}{A}\right)^{2}
&=&8\pi p-A^{4}(B_{0})^{2}
\eea
It is easy to show that the class of cosmological 
models of Appenix A, satisfies identically 
Eqs. (\ref{einback}). Now since we have 
$\nabla ^{\nu }(G_{\mu \nu }^{(0)}+\delta G_{\mu \nu })
=0+{\cal O}(h^{3})$ we will also have from Eq. (\ref{expan}) 
$\nabla ^{\nu }(T_{\mu \nu }^{(0)}+\delta T_{\mu \nu })
=0+{\cal O}(h^{3})$. So these conservation equations need 
not be considered separately, because they are embodied in 
Eqs. (\ref{einback}) and in
\bea 
\label{einpert}
\delta G_{\mu \nu }=8\pi \delta T_{\mu \nu }. 
\eea
The content of Eqs. (\ref{einpert}) determines the fluid and 
EM-field perturbations in terms of the propagating GW and 
will be given below.

The only remaining set of equations that need to be considered 
is Maxwell's equations $\nabla _{\nu }F^{\mu \nu }=4\pi J^{\mu }$ 
and $F_{[\alpha \beta ;\gamma ]}=0$. Here the current density is 
given by\\ 
$J^{\mu }=\rho _{Q}u^{\mu }+\sigma F^{\mu \nu }u_{\nu }$. The 
local charge density $\rho _{Q}=Z_{e}n_{i}-en_{e}\simeq 0$, 
i.e., can be taken as zero due to the mobility of the lighter 
electron with respect to the other ion species, or due to  
exact cancelation (i.e., in the case for example of an 
electron-positron plasma). Also we have 
$\sigma =1/(4\pi \eta )$ for the finite conductivity 
of the plasma fluid, with $\eta $ the resistivity. 

Now from the $(\alpha \beta ;\gamma )=(0xy)$ and 
$(xyz)$-components of the sourceless Maxwell's equations 
we exactly obtain Eq. (\ref{backmag}) of Appendix A, and 
\bea 
\label{bez}
\delta B^{z}(t,z)=\frac{H_{0}}{A^{8}}(h_{+}^{2}+h_{\times }^{2}),
\eea
while from the $(0xz)$ and $(0yz)-$components we obtain 
respectively 
\bea 
\label{maxmagn}
(A^{2}W^{2}\delta B^{y})_{,t}+A^{2}(\delta E^{x})_{,z}&=&0
\nn \\
(A^{2}W^{2}\delta B^{x})_{,t}-A^{2}(\delta E^{y})_{,z}&=&0
\eea
From the sourcefull Maxwell's equations we obtain that\\ 
$\delta E^{z}=0$ and 
\bea 
\label{maxelec}
(\delta E^{x})_{,t}+(\delta B^{y})_{,z}+(\delta E^{x})
\left[2\frac{\dot{A}}{A}+\frac{\dot{W}}{W}-\right.\nn \\
-\frac{h_{+}\dot{h}_{+}+h_{\times }\dot{h}_{\times }}{A^{4}}
\left.+2\frac{\dot{A}}{A}\frac{(h_{+}^{2}+h_{\times }^{2})}{A^{4}}\right]
-\nn \\-(\delta B^{y})
\left[\frac{h_{+}\dot{h}_{+}+h_{\times }\dot{h}_{\times }}{A^{4}}\right]
+\nn \\+4\pi \sigma [\delta E^{x}+B_{0}A^{2}\delta u^{y}]=0
\eea
along with a similar equation, that results from the substitutions 
$\delta E^{x}\rightarrow \delta E^{y}$, $\delta B^{y}
\rightarrow -\delta B^{x}$ and $\delta u^{y}\rightarrow -\delta u^{x}$. 
These are the propagation equations for the EM-field perturbations. 

We now consider Eqs. (\ref{einpert}). From the $(0x)$, $(0y)$ and 
$(0z)$ components we obtain the fluid's four-velocity perturbations 
\bea 
\label{velpert}
\delta u^{x}&=&-\frac{1}{4\pi }\frac{A^{2}B_{0}}{(\rho +p)}\delta E^{y}
\nn \\
\delta u^{y}&=&\frac{1}{4\pi }\frac{A^{2}B_{0}}{(\rho +p)}\delta E^{x}
\eea
and 
\bea 
\label{uzpert}
8\pi (\rho +p)\delta u^{z}&=&\frac{1}{2A^{4}}(\dot{h}_{+}h_{+}^{'}
+\dot{h}_{\times }h_{\times }^{'})+\nn \\&+&\frac{1}{A^{4}}
(h_{+}\dot{h}_{+}^{'}+h_{\times }\dot{h}_{\times }^{'})-\nn \\
&-&\left(\frac{\dot{A}}{A}+\frac{\dot{W}}{W}\right)
\frac{(h_{+}h_{+}^{'}+h_{\times }h_{\times }^{'})}{A^{4}}
\eea

From the $(00)$ component we obtain 
\bea 
\label{denspert}
8\pi \delta \rho &=&\left(\frac{\dot{A}}{A}\right)^{2}
\frac{(h_{+}^{2}+h_{\times }^{2})}{A^{4}}-\nn \\
&-&\frac{1}{4}
\frac{[(\dot{h}_{+})^{2}+(\dot{h}_{\times })^{2}]}{A^{4}}
+\nn \\&+&2\left(\frac{\dot{A}}{A}\frac{\dot{W}}{W}\right)
\frac{(h_{+}^{2}+h_{\times }^{2})}{A^{4}}+\nn \\
&+&\frac{3}{4W^{2}}
\frac{[(h_{+}^{'})^{2}+(h_{\times }^{'})^{2}]}{A^{4}}-
\nn \\ &-&\left(\frac{\dot{W}}{W}\right)
\frac{(h_{+}\dot{h}_{+}+h_{\times }\dot{h}_{\times })}{A^{4}}+
\nn \\&+&\frac{1}{W^{2}A^{4}}[h_{+}h_{+}^{''}+h_{\times }h_{\times }^{''}]
-\nn \\&-&2\frac{H_{0}B_{0}}{A^{4}}[h_{+}^{2}+h_{\times }^{2}]
\eea
From proper combinations of the $(xx)$, $(yy)$ and $(zz)$ components 
of Eqs. (\ref{einpert}) we obtain the pressure perturbation
\bea
\label{pressu} 
8\pi \delta p&=&-\frac{1}{4W^{2}}
\frac{[(h_{+}^{'})^{2}+(h_{\times }^{'})^{2}]}{A^{4}}+\nn \\&+&
\left[-2\frac{\ddot{A}}{A}+3\left(\frac{\dot{A}}{A}\right)^{2}\right]
\frac{(h_{+}^{2}+h_{\times }^{2})}{A^{4}}-\nn \\&-&
4\left(\frac{\dot{A}}{A}\right)
\frac{(h_{+}\dot{h}_{+}+h_{\times }\dot{h}_{\times })}{A^{4}}
+\frac{3}{4}\frac{[(\dot{h}_{+})^{2}+(\dot{h}_{\times })^{2}]}{A^{4}}+
\nn \\&+&
\frac{(h_{+}\ddot{h}_{+}+h_{\times }\ddot{h}_{\times })}{A^{4}}
+2\frac{H_{0}B_{0}}{A^{4}}[h_{+}^{2}+h_{\times }^{2}],
\eea
the constraint 
\bea 
\label{constr}
-\frac{1}{2}\frac{[(\dot{h}_{+})^{2}+(\dot{h}_{\times })^{2}]}{A^{4}}
+\left(\frac{\dot{A}}{A}+\frac{\dot{W}}{W}\right)
\frac{(h_{+}\dot{h}_{+}+h_{\times }\dot{h}_{\times })}{A^{4}}-\nn \\
-\frac{1}{W^{2}A^{4}}[h_{+}h_{+}^{''}+h_{\times }h_{\times }^{''}]
-\frac{4H_{0}B_{0}}{A^{4}}[h_{+}^{2}+h_{\times }^{2}]=0\nn \\
\eea
and the propagation equation for the {\it cross} 
polarization of the gravitational wave
\bea 
\label{crosspol}
\Box h_{+}&:=&\frac{1}{2}\left(\ddot{h}_{+}-\frac{1}{W^{2}}h_{+}^{''}\right)
+\nn \\&+& 2\left(\frac{\dot{A}}{A}\right)^{2}h_{+}-
\left(\frac{\dot{A}}{A}\right)\dot{h}_{+}+
\frac{1}{2}\left(\frac{\dot{W}}{W}\right)\dot{h}_{+}-\nn \\
&-&\left(\frac{\ddot{W}}{W}+2\frac{\ddot{A}}{A}+\left(\frac{\dot{A}}{A}\right)^{2}\right.
\left.+2\frac{\dot{A}}{A}\frac{\dot{W}}{W}\right)h_{+}=\nn \\
&=&[8\pi p+A^{4}(B_{0})^{2}]h_{+}
\eea
Finally from the $(xy)$ component we obtain the propagation equation 
for the {\it times} polarization of the GW\\
$\Box h_{\times }=[8\pi p+A^{4}(B_{0})^{2}]h_{\times }$. 
Eqs. (\ref{bez})-(\ref{crosspol}) constitute a closed system 
of equations that we study below. 

\section{Analytical Solutions}
Substituting the solution of Appendix A into Eq. (\ref{crosspol}) we 
obtain (setting also $h_{+}(t,z)=e^{ik_{g}z}h_{+}(t)$)
\bea 
\label{fulleq}
\ddot{h}_{+}+\frac{k_{g}^{2}}{t^{2l}}h_{+}-
\left(\frac{2\gamma }{1+\gamma }\right)\frac{\dot{h}_{+}}{t}+
\left(\frac{2\gamma }{1+\gamma }\right)\frac{h_{+}}{t^{2}}=0
\eea
A 
similar equation is satisfied by the other polarization. Also 
the constraint equation, Eq. (\ref{constr}) is satisfied if 
both polarizations obey 
\bea 
\label{constr1}
(\dot{h}_{+})^{2}-\frac{(3-\gamma )}{(1+\gamma )}
\frac{h_{+}\dot{h}_{+}}{t}-\frac{k_{g}^{2}}{t^{2l}}
h_{+}^{2}+
\frac{(1-\gamma )(3\gamma -1)}{(1+\gamma )^{2}t^{2}}
h_{+}^{2}=0\nn \\
\eea

We can take as final time of our analysis any time before 
the {\it recombination} epoch 
$t_{fin}\simeq 10^{13}sec$, where presumably the cosmological plasma 
ceases to exist. Also the initial time can be 
any time after the typical time that 
inflation starts ($t_{0}\simeq 10^{-34}sec$), or the end of the reheating 
period ($t_{0}\simeq 10^{-31}sec$). Between these we recognize two eras:  
The first is the {\it small}-time era where in Eq. (\ref{crosspol}) the 
{\it second} term is small compared with the next two. The second is the 
{\it large}-time era where the {\it last} term can be omitted. Equating these 
we have an estimate of this transition time as 
\bea 
\label{transtime}
t_{*}=\left[\frac{\gamma }{(1+\gamma )k_{g}^{2}}\right]
^{1/2(1-l)}
\eea
We observe that this time scale coincides with the time that 
the mode, with {\it comoving} wavevector $k_{g}=1/\lambda _{c}$, 
enters the Horizon.  Indeed the {\it physical} wavelength\\ 
$\lambda _{phys}=W\lambda _{c}$ must be smaller than the 
Horizon scale, $\lambda _{phys}\leq l_{H}=c/H_{W}=cW/\dot{W}$. 
This gives the estimate $t_{*}^{(H)}=[l^{2}/k_{g}^{2}]^{1/2(1-l)}$, 
which is of the same order with\\ Eq. (\ref{transtime}). 

\begin{figure}
\resizebox{\hsize}{!}{\includegraphics{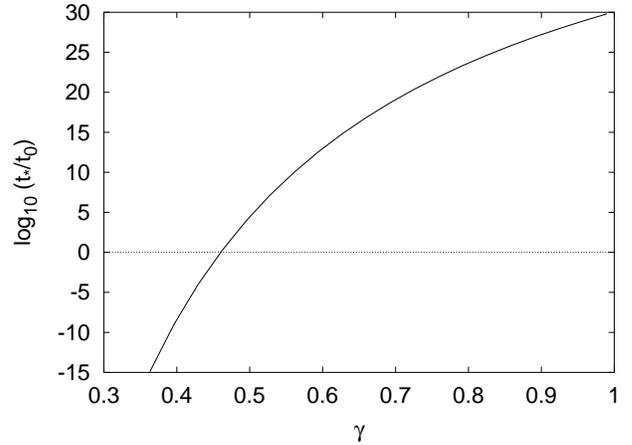}}
\caption{The transition time $t_{*}$, of Eq. (\ref{transtime}), 
as a function of the parameter $\gamma $}
\label{transtf}
\end{figure}

We identify the modes by their wavenumber $k_{g}$ and their 
initial-time frequency $f_{0}$ so that 
$2\pi f_{0}=k_{g},$ (we take throughout $c=1)$. At the end of the 
time interval $t_{fin}$ let the frequency be $f_{fin}=f_{m}(KHz)$. 
Due to the cosmological redshift we have $f_{0}=f_{fin}W(t_{fin})/W(t_{0})$. 
If we denote $u:=(t/t_{0})$ then the transition time $t_{*}$ of 
Eq. (\ref{transtime}) is given by 
\bea 
\label{transtime1}
log_{10}(u_{*})&=&\frac{1}{2(1-l)}
\left[(62-162l)+\right. \nn \\
&+&\left.log_{10}\left(\frac{\gamma }{4\pi ^{2}(1+\gamma )f_{m}^{2}}\right)\right]
\eea
and is plotted in Fig. (\ref{transtf}). It is evident that this can 
be obtained for other boundary values for $t_{0},\; t_{fin}$.

\subsection{Small-t Solutions}

There are many types of solutions, for example power-law 
solutions, namely when the {\it second} term on 
the l.h.s of Eq. (\ref{fulleq}) is omitted. The first is 
\bea 
\label{fstsmall}
h_{+}^{(1)}(t)=C_{1}t+C_{2}t^{2\gamma /(1+\gamma )}
\eea
and it is evident that by substituting into Eq. (\ref{fulleq}) 
the error is of order ${\cal O}(1/t^{2l})$. We also find that 
Eq. (\ref{constr1}) is satisfied with remaining error terms of 
order ${\cal O}(1/t^{2l})$, provided that $C_{1}=0$. 

The second solution we found is 
\bea 
\label{sndsmall}
h_{+}^{(2)}(t)=C_{3}t^{1/2}exp
\left[ik_{g}\int_{t_{0}}^{t}\frac{du}{u^{l}}\right]
\eea
which, when substituted into Eq. (\ref{fulleq}) gives error terms 
of order ${\cal O}(1/t^{2})$! Also the constraint equation 
Eq. (\ref{constr1}) is satisfied to the order ${\cal O}(1/t^{l+1})$. 
Here the mode is outside the 
Horizon and the amplitude of the GW is assumed to be driven by 
the so-called cosmological pump field. However, for our purposes 
to show that certain classes of gravitational waves can induce 
strong electric field perturbations, 
we use the following solution
\bea 
\label{thrdsmall}
h_{+}(t)=t^{2\gamma /(1+\gamma )}exp
\left[bt^{l}\right]
\eea
with $b$ a constant. Upon substituting into Eq. (\ref{fulleq}) 
we find errors of order ${\cal O}(1/t^{2(1-l)})$ and of the same 
order for Eq. (\ref{constr1}).

In order to show the consistency of the whole model we consider 
Eq. (\ref{fulleq}) without the first term. This can be solved analytically 
to obtain the {\it exact} solution 
\bea 
\label{frthsmall}
h_{+}(t)&=&h_{0}texp
\left[\frac{k_{g}^{2}(1+\gamma )^{2}}{8\gamma ^{2}}t^{\frac{4\gamma }{(1+\gamma )}}\right]
=\nn \\&=&h_{0}texp
\left[\frac{(1+\gamma )}{8\gamma }(t/t_{*})^{4\gamma /(1+\gamma )}\right]
\eea
Computing the second derivative we obtain 
\bea 
\label{doublder}
\ddot{h}_{+}&=&\dot{h}_{+}
\left[\frac{1}{t}+\frac{k_{g}^{2}(1+\gamma )}{2\gamma }t^{(3\gamma -1)/(1+\gamma )}\right]
+\nn \\&+&h_{+}
\left[-\frac{1}{t^{2}}+\frac{k_{g}^{2}(3\gamma -1)}{2\gamma }\frac{1}{t^{2l}}\right]
\eea
Thus the seccond derivative will be small with respect to the first 
and zeroth order derivatives, as times passes, if the second term, in the 
first set of brackets, is smaller than the first term. This {\it exactly} 
reproduces the condition of Eq. (\ref{transtime}).

\subsection{Large-t Solution}

Here we assume that the last term on the l.h.s of 
Eq. (\ref{fulleq}) is omitted and we have the usual 
solution of a mode that is inside the cosmological Horizon 
and its amplitude decreases with time,
\bea 
\label{large}
h_{+}(t)=\frac{1}{t^{2l}}
\left(D_{1}e^{-ik_{g}t}+D_{2}e^{ik_{g}t}\right)
\eea
When substituted into Eq. (\ref{fulleq}) we have error terms 
of the order ${\cal O}(1/t^{2l+1})$. Also we find that 
Eq. (\ref{constr1}) is satisfied to the order 
${\cal O}(1/t^{4l})$.

We can match the two solutions of 
Eqs. (\ref{frthsmall}) and (\ref{large}) at $t=t_{*}$ 
with continuity of the first derivatives and we obtain
\bea 
\label{match}
e^{ik_{g}t_{*}}D_{2}&=&
\frac{(3/2+2l+ik_{g}t_{*})}{2ik_{g}t_{*}}h_{0}
(t_{*})^{2l+1}exp\left(\frac{1+\gamma }{8\gamma }\right)\nn \\
e^{-ik_{g}t_{*}}D_{1}&=&
\frac{(-3/2-2l+ik_{g}t_{*})}{2ik_{g}t_{*}}h_{0}
(t_{*})^{2l+1}exp\left(\frac{1+\gamma }{8\gamma }\right)
\nn \\
\eea
This solution when substituted into Eq. (\ref{denspert}) 
gives the usual oscillating behaviour of the density 
perturbations when the driver (GW) enters the Horizon, 
after the time $t=t_{*}$.

\section{Numerical Results\label{NumRes}}

\begin{figure}
\resizebox{\hsize}{!}{\includegraphics{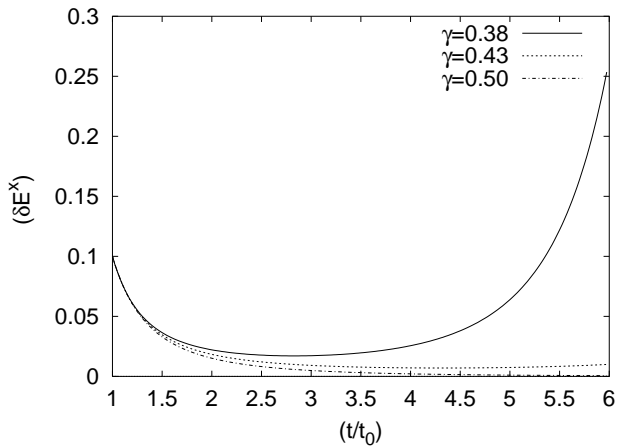}}
\caption{Electric field perturbation as a function of $\gamma $}
\label{figur1}
\end{figure}

\begin{figure}
\resizebox{\hsize}{!}{\includegraphics{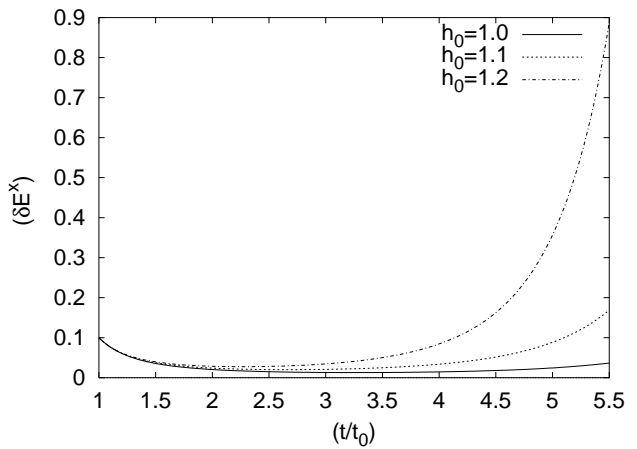}}
\caption{Electric field perturbation as a function of $h_{0}$}
\label{figur2}
\end{figure}

We consider Eq. (\ref{maxelec}) and the first of 
Eqs. (\ref{maxmagn}), where we take the magnetic field 
perturbations equal to zero for simplicity. We obtain 
\bea 
\label{elnumer}
(\delta E^{x})_{,u}&+&(\delta E^{x})
\left[\frac{2}{(1+\gamma )u}-\frac{2h_{+}(h_{+})_{,u}}{u^{2}}\right.
+\frac{2h_{+}^{2}}{u^{3}}+\nn \\&+&
4\pi \sigma _{0}
\left.\left(1+\frac{(1-\gamma )(3\gamma -1)}{(1+\gamma )(3-\gamma )}\right)\right]
=0
\eea
where the conductivity is in units of $(1/t_{0})$. It is evident that 
the conductivity counteracts the electric field perturbations 
that arise from the passage of the GW, whereas the constant $b$ in 
Eq. (\ref{thrdsmall}) acts in favor of steeper in time electric field 
perturbations. For an initial value of $(\delta E^{x})(t_{0})=0.1$ 
we integrate numerically Eq. (\ref{elnumer}) for $\sigma _{0}=0.1$, 
$h_{0}=1.0$ and $b=1$ and study the dependence of the electic field 
perturbation on the parameter $\gamma $. This is shown in 
Fig. (\ref{figur1}). Also for $\sigma _{0}=0.1$, $b=1$ and $\gamma =0.4$ 
we study the dependence of the electic field perturbation on the parameter 
$h_{0}$, effectively the (normalized) amplitude of the metric 
perturbations. This behaviour is shown in Fig. (\ref{figur2}).

\section{Discussion}

We have considered the propagation of a gravitational 
wave perturbation parallel to the ambient magnetic field of an  
anisotropic magnetized cosmological 
spacetime, non-linearly coupled with its temporal derivatives. It 
is known that in the first order (linear) approximation 
with respect to the amplitude of the perturbation, no 
electric field or density perturbations arise. However in 
the non-linear level this behaviour changes. Namely while 
certain classes of power-law solutions, for these metric 
perturbations, {\it cannot} 
induce increasing in time electric field perturbations, certain 
other classes of solutions can do so. 

Our numerical results and their 
theoretical counterparts show that while power-law solutions, such as 
those of Eqs. (\ref{fstsmall}) and (\ref{sndsmall}), are not capable 
of inducing increasing in time, electric field perturbations, solutions 
of the form of Eqs. (\ref{thrdsmall}) and (\ref{frthsmall}) can do so. 
Moreover these solutions are consistent with the {\it small-t} and 
{\it large-t} assumptions, as these are encoded in Eq. (\ref{transtime}). 
The consistency of the whole theoretical model is further shown 
by the numerical treatment of Eq. (\ref{maxelec}). The electric 
field perturbations depend on four parameters: First on the conductivity 
$\sigma _{0}$, which in an evident and expected manner counteracts 
the generation of electric field perturbations. Second on the 
constant $b$, in the class of solutions of Eq. (\ref{thrdsmall}) 
which again in an expected manner acts in favour of steeper in 
time electric field and density perturbations. The (normalized) 
amplitude $h_{0}$ of the metric perturbations is the third parameter, 
as it is shown in Fig. (\ref{figur2}) and $\gamma $ is the fourth 
parameter, as it is shown in Fig. (\ref{figur1}). Since the 
Hubble parameter in the $z-$direction is 
$H_{z}:=(\frac{\dot{W}}{W})=(1-\gamma )/(1+\gamma )t$ we see that 
higher values of the $\gamma -$parameter imply smaller values 
of this parameter, which in turn acts in favour of the electric 
field perturbations. 

Our results hopefully shed some light on the full non-linear 
behaviour of conductive plasmas, in anisotropic cosmological 
models, coupled with an ambient magnetic field and driven by 
(not necessarily weak) metric perturbations.

\begin{acknowledgments}
This work was supported by the Greek Ministry of Education through the
PYTHAGORAS research program.
\end{acknowledgments}
\section*{Appendix A }

We present here for convenience the class of anisotropic 
magnetized cosmologies, with perfect fluid content, known as 
Thorne's models \cite{thorn}. 
The scale factors are given by $A(t)=t^{1/2}$ 
and $W(t)=t^{l}$ as used in Eq. (\ref{metr}). Here 
\bea 
l:=\frac{1-\gamma }{1+\gamma },\; \; \; (\frac{1}{3}<\gamma \leq 1)
\eea
The matter energy density and pressure are given by 
\bea 
\rho (t)=\frac{3-\gamma }{16\pi (1+\gamma )^{2}t^{2}},
\; \; \; p(t)=\gamma \rho (t)
\eea
Finally the magnetic field points in the $z-$direction and 
is given by 
\bea 
\label{backmag}
B_{0}(t)=\frac{H_{0}}{t^{2}},\; \; \; 
H_{0}:=\frac{(1-\gamma )^{1/2}(3\gamma -1)^{1/2}}{2(1+\gamma )}
\eea
Also we note that the relativistic Alfven velocity is \\
$u_{A}^{2}=v_{A}^{2}/(1+v_{A}^{2})$ where in conformity 
with the notation used in \cite{thorn} we have 
\bea 
v_{A}^{2}:=\frac{(F^{x}_{y})^{2}}{4\pi \rho }=
\frac{(1-\gamma )(3\gamma -1)}{(3-\gamma )}
\eea
\section*{Appendix B }

We present here two of the components of Einstein's tensor, 
just for the purpose of displaying the relative complexity 
of the task, computed from Eq. (\ref{metr}), after they have 
been carefully expanded up to the {\it second} order with respect to 
the GW amplitude.  This is a difficult task and has 
been performed with great care. We obtain 
\bea
\label{g00}
G_{00}
&=&\left(\frac{\dot{A}}{A}\right)^{2}+
2\left(\frac{\dot{A}\dot{W}}{AW}\right)+
\left(\frac{\dot{A}}{A}\right)^{2}
\left[\frac{h_{+}^{2}}{A^{4}}+\frac{h_{\times }^{2}}{A^{4}}\right]\nn \\
&-&\frac{1}{4}\left[\frac{(\dot{h_{+}})^{2}}{A^{4}}+\frac{(\dot{h_{\times }})^{2}}{A^{4}}\right]
\nn \\
&+&2\left(\frac{\dot{A}\dot{W}}{AW}\right)
\left[\frac{h_{+}^{2}}{A^{4}}+\frac{h_{\times }^{2}}{A^{4}}\right]\nn \\
&+&\frac{3}{4W^{2}}
\left[\frac{(h_{+}^{'})^{2}}{A^{4}}+\frac{(h_{\times }^{'})^{2}}{A^{4}}\right]
\nn \\&-&\left(\frac{\dot{W}}{W}\right)
\left[\frac{h_{+}\dot{h}_{+}}{A^{4}}+\frac{h_{\times }\dot{h}_{\times }}{A^{4}}\right]
\nn \\
&+&\frac{1}{W^{2}A^{4}}[h_{+}h_{+}^{''}+h_{\times }h_{\times }^{''}]
\eea
and
\bea 
\label{gxx}
G_{xx}&=&A^{2}
\left[-\frac{\ddot{A}}{A}-\frac{\ddot{W}}{W}-\frac{\dot{A}\dot{W}}{AW}\right]
+\frac{1}{2}\left(\ddot{h}_{+}-\frac{1}{W^{2}}h_{+}^{''}\right)+\nn \\
&+&2\frac{(\dot{A})^{2}}{A^{2}}h_{+}-\left(\frac{\dot{A}}{A}\right)\dot{h}_{+}
+\frac{1}{2}\frac{\dot{W}}{W}\dot{h}_{+}-\nn \\
&-&\left(\frac{\ddot{W}}{W}+2\frac{\ddot{A}}{A}+\left(\frac{\dot{A}}{A}\right)^{2}+\right.
\left.2\frac{\dot{A}\dot{W}}{AW}\right)h_{+}+\nn \\
&+&\frac{1}{4}\left[\frac{(\dot{h}_{+})^{2}+(\dot{h}_{\times })^{2}}{A^{2}}\right]
-3\left(\frac{\dot{A}}{A}\right)\frac{h_{+}\dot{h}_{+}+h_{\times }\dot{h}_{\times }}{A^{2}}
+\nn \\&+&
\left(\frac{\dot{W}}{W}\right)\frac{h_{+}\dot{h}_{+}+h_{\times }\dot{h}_{\times }}{A^{2}}-
\frac{1}{4W^{2}}
\left[\frac{(h_{+}^{'})^{2}+(h_{\times }^{'})^{2}}{A^{2}}\right]+\nn \\&+&
\left[-2\left(\frac{\ddot{A}}{A}\right)-2\frac{\dot{A}\dot{W}}{AW}+3\left(\frac{\dot{A}}{A}\right)^{2}\right]
\frac{h_{+}^{2}+h_{\times }^{2}}{A^{2}}\nn \\
&+&\frac{h_{+}}{A^{2}}\left[\ddot{h}_{+}-\frac{h_{+}^{''}}{W^{2}}\right]+
\frac{h_{\times }}{A^{2}}\left[\ddot{h}_{\times }-\frac{h_{\times }^{''}}{W^{2}}\right]
\eea
Here a dot denotes derivative with respect to the time $t$, 
while a prime denotes derivative with respect to the spatial variable 
$z$. Similar expressions are obtained for the rest of the 
non-zero components, namely for $G_{yy}$, $G_{zz}$, 
$G_{xy}$ and $G_{0z}$.


\end{document}